# Interface-Induced Ordering of Gas Molecules Confined in a Small Space

Ing-Shouh Hwang*, Yi-Hsien Lu, Chih-Wen Yang, Chung-Kai Fang, and Hsien-Chen Ko

Institute of Physics, Academia Sinica, Nankang, Taipei 115, Taiwan, R.O.C.

*Correspondence to:  ishwang@phys.sinica.edu.tw

## Abstract

The thermodynamic properties of gases have been understood primarily through phase diagrams of bulk gases.  However, observations of gases confined in a nanometer space have posed a challenge to the principles of classical thermodynamics.  Here, we investigated interfacial structures comprising either $O_2$ or $N_2$ between water and a hydrophobic solid surface by using advanced atomic force microscopy techniques.  Ordered epitaxial layers and cap-shaped nanostructures were observed.  In addition, pancake-shaped disordered layers that had grown on top of the epitaxial base layers were observed in oxygen-supersaturated water.  We propose that hydrophobic solid surfaces provide low-chemical-potential sites at which gas molecules dissolved in water can be adsorbed.  The structures are further stabilized by interfacial water.  Gas molecules can agglomerate into a condensed form when confined in a sufficiently small space under ambient conditions.  The ordering and thermodynamic properties of the confined gases are determined primarily according to interfacial interactions.  The crystalline solid surface may even induce a solid-gas state.

Gases exist throughout the universe and are essential in daily life as well as science and technology.   Gases are generally defined as molecules that have boiling points below room temperature, such as the small nonpolar molecules $N_2$, $O_2$, He, and Xe.  Gases are vapor under ambient conditions because van der Waals (VDW) interactions among gas molecules are much weaker than thermal energy.  Condensing gas molecules into a liquid or solid state, based on the phase diagrams of bulk gases, requires high pressures or cryogenic techniques.  However, numerous puzzling observations regarding gases confined in a small space have been reported.  For example, gases have been observed to accumulate in a cap-shaped space on a nanometer scale at solid-water interfaces, mainly hydrophobic-water interfaces, under ambient conditions (*1-13*).  The cap-shaped structures are generally considered interfacial nanobubbles (INBs) that feature gas molecules in their vapor (gaseous) phase.  However, their thermodynamic stability, nature, nucleation, and other properties and behaviors remain unclear.  Theoretical prediction has indicated that gases inside a bubble of nanometer size should dissolve into the surrounding water in a short time (*14*) because of its high internal pressure, $P_{in}$, which can be described using the Young-Laplace equation,

Laplace pressure   $P_L = P_{in} - P_0 = 2\Gamma / r$     (1)

where $P_0$ is the liquid pressure (approximately 1 atm in most laboratory conditions), $\Gamma$ is the surface tension of the interface between liquid and gas, and $r$ is the radius of the bubble (**Supplementary text S1**).  Numerous atomic force microscopy (AFM) observations have shown that INBs are stable for hours or days (*1-12*), which are at least 10-11 orders of magnitude longer than the theoretical lifetime estimated based on the Young-Laplace equation (*9*).  Another notable observation is quasi-two-dimensional (quasi-2D) micropancakes, which are layered structure of gas with a thickness of 0.3-10 nm and a width of sub-micron or micron size, on certain hydrophobic surfaces in air-supersaturated water (*15-18*). These structures are as stable as INBs and occasionally INBs can be located on top of micropancakes. The nature of micropancakes remains unclear (**Supplementary text S2**).

Another example is the observation of liquid or solid nanoprecipitates of inert gases inside solids at room temperature (*19-35*). These nanoprecipitates remain stable even at temperatures at which the precipitates can change shape and position or undergo coalescence. Compressive high pressures were expected for the condensation of inert gases. Generally, highly pressurized gases are thermodynamically unfavorable. Thus, it is unclear why these precipitates exhibit a high stability.

In this work, we conducted AFM studies of the interface between water and a hydrophobic substrate, highly ordered pyrolytic graphite (HOPG), in water supersaturated with either oxygen or nitrogen. These two gases were chosen because they are the two major components of air. Both gases exhibited ordered epitaxial base layers and cap-shaped soft nanostructures resembling INBs. The latter were probably liquid-like gas agglomerates at the HOPG-water interface. Regarding oxygen, structures resembling the micropancakes were observed and they exhibited disordered dense gas layers on top of the ordered base layers. We conclude that the ordering of nitrogen and oxygen inside these interfacial structures is induced by the crystalline HOPG substrate and further stabilized by the interfacial water on the top. These interfacial structures comprising gas molecules may possess novel properties that can explain several scientific problems. Understanding the underlying mechanisms may enable numerous technological applications. In addition, the interface-induced ordering can be generalized to understand previous observation of the solid or liquid nanoprecipitates of inert gases inside solids and other phenomena.

## *AFM study of the structures of oxygen and nitrogen at the HOPG-water interface*

After oxygen-supersaturated water was deposited onto a clean HOPG sample, we observed considerably complex interfacial structures. A typical case is shown in Figs. 1, 2, and **S1-S4**. We used two advanced AFM techniques, the frequency-modulation (FM) and PeakForce (PF) modes, because each has distinct advantages in probing the interfacial structures. Here, $t$ is defined as 0 when the water is deposited onto the HOPG surface. Only the late stage of growth is shown because the interface was highly dynamic following water deposition.

Figure 1 shows FM-AFM images of the interfacial structures. Cap-shaped nanostructures were present and tended to be located on top of thick pancake-shaped layers (Figs. 1A and 1B). The thicknesses of the quasi-2D structures varied among areas (Fig. 1A). High-resolution imaging indicated no bare HOPG region (Figs. 1B to 1C). The lower (darker) regions were the ordered base layers, which featured thicknesses varying from one to four monolayers and comprised domains of two sets of row-like patterns. The ordered structures can be discerned easily in the dissipation image (Fig. 1D). Bright (high dissipation) regions exhibited row-like patterns with row separation of 4-5 nm, and dark (low dissipation) regions are featureless, suggesting that the pancake- and cap-shaped structures were disordered. In Fig. 1C, a few small disordered regions were only approximately 0.5 nm higher than the neighboring ordered structure, suggesting that one disordered molecular layer was located on top of the ordered base layers. Most thick pancake-shaped layers were 2-4 nm higher than the base layers (Fig. S1), and their shape and thickness resembled those of the micropancakes reported in previous AFM studies (*15-18*). Investigating various HOPG–water interfaces indicated that the thick pancake-shaped structures appear only when water is supersaturated with air or oxygen. These structures have never been observed when pre-degassed water or water saturated or supersaturated with pure nitrogen is used. This strongly suggests that micropancakes form through the precipitation of oxygen molecules at the interface.

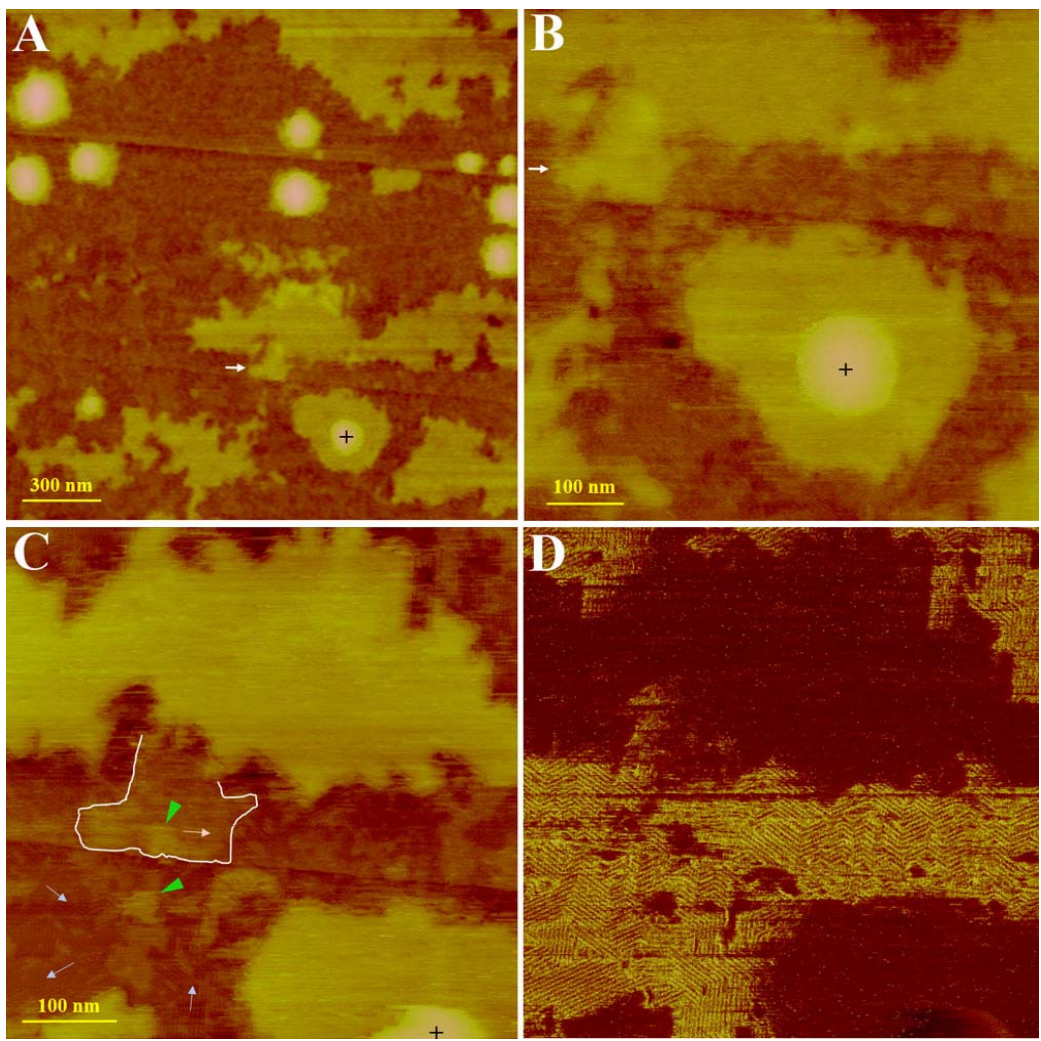

**Fig. 1. FM-AFM measurements of a HOPG-water interface after oxygen-supersaturated water was deposited.** (A) Topographic image acquired at t=300 min. A cap-shaped nanostructure marked with "＋"serves as a marker for comparison with other images. (B) Higher-resolution height image of a region around the cap-shaped nanostructure marked with "＋" acquired at t=310 min. Because of the large height variation of various structures, the contrast of the row-like patterns in the base layers cannot be clearly shown. (C) Topographic image of a similar area acquired at t=205 min. The region outlined by white lines was later covered by a thick disordered layer, as indicated by the white arrow in (A) and (B). In the ordered base layers, two sets of row-like structures existed; one set was oriented along a zigzag direction (indicated using pink arrows) of the HOPG substrate and the second was oriented along an arm-chair direction (indicated using blue arrows). The green arrow indicates areas with a disordered molecular layer. (D) Dissipation image acquired simultaneously with the image shown in (C). We note that primarily flat terraces and substrate step edges were observed when pre-degassed water was deposited on clean HOPG (*36-38*). Domains of monolayer row-like patterns might appear, but the coverage was typically less than 10% of the interface after 8 hours of water deposition. Thus the 2D and 3D patterns shown here were produced by oxygen accumulating at the interface.

Domains of row-like patterns were observed in regions originally exhibiting thick disordered pancake-shaped structures when we switched to the PF mode to study the interface. An example is shown in Fig. 2A, S2 and S3. These patterns were similar to the surrounding ordered base layers, although their heights were approximately one molecular layer lower. The cap-shaped nanostructures could still be seen, but their apparent heights were significantly smaller than those measured with the FM mode (Figs. S1-S3). Our measurements of force versus tip-sample separation, or force curves, indicate a snap-in when the tip touched a cap-shaped nanostructure or a thick pancake-shaped layer (Fig. S4). The tip penetrated into the structures to a certain depth, rather than traced their interfaces with water, to offset the attractive snap-in force. For the same reason, the tip revealed domains of the stiff ordered structures under the pancake-shaped disordered layer. Growth of a thick disordered layer on top of the ordered base layers was observed in the region outlined by a white line in Fig. 1C compared with subsequently acquired images shown in Figs. 1A and 1B. We probed several areas on this sample and observed ordered base layers beneath all disordered quasi-2D structures. Similar structures were also observed for micropancakes at the HOPG-water interface when the water was supersaturated with air.

The apparent height of the cap-shaped nanostructure decreased as the peak force increased (Figs. 2 A-D) because the tip penetrated deeper into the structure. When the peak force was increased to +650 pN, ordered row-like structures under the cap-shaped nanostructure became visible (Fig. 2C). When the peak force was increased to +1250 pN (Fig. 2D), the cap-shaped nanostructure was not observed and numerous regions of the ordered structures were seemingly destroyed by the high force and became fuzzy, particularly the region under the thick pancake-shaped layer. When the peak force was returned to +100 pN (Fig. 2E), the cap-shaped nanostructure appeared again, but its apparent width and height substantially decreased compared with those shown in Fig. 2A. In addition, the structures in numerous areas of the base layers remained fuzzy, indicating disordering in those areas after imaging with the high peak force. After continuous imaging using a low peak force (Fig. 2F), ordered row-like structures gradually appear in the fuzzy regions over the time scale of minutes or longer.

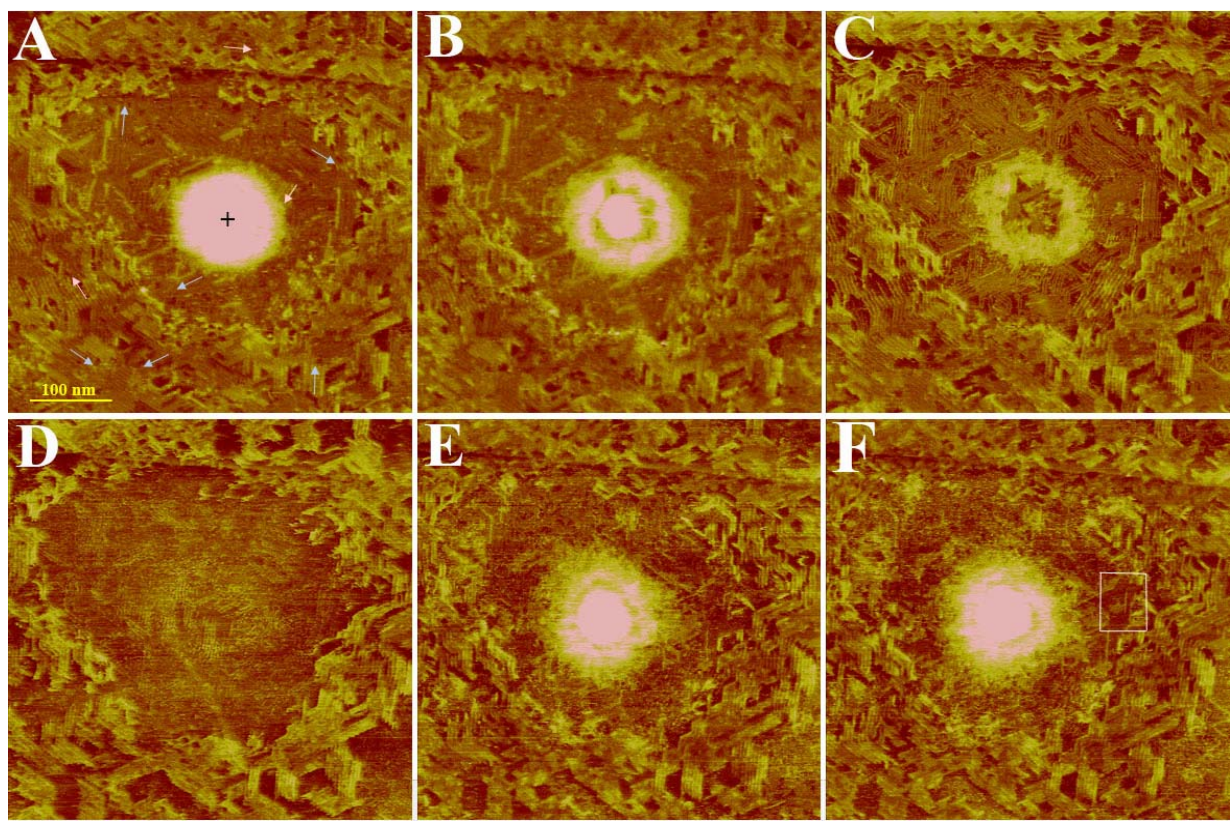

**Fig. 2. Height images of a HOPG-water interface acquired with PF mode after oxygen-supersaturated water was deposited.** Peak forces of +100 pN (A), +250 pN (B), +650 pN (C), +1250 pN (D) were applied. The peak force was returned to 100 pN in (E) and (F). In the region corresponding to the thick pancake-shaped layer, the tip penetrated through the disordered layer to image the underlying ordered base layers, which appeared approximately 0.5 nm lower than the surrounding ordered layers. Ordered structures, comprising two sets of three rotated row-like structures, were observed. One set with the rows oriented along the arm-chair directions is indicated using blue arrows and the second set, which was oriented along the zig-zag directions, is indicated with pink arrows. The outlined region in (F) indicates an area showing formation of new ordered structures.

As shown in Figs. S2B-S2D, the thick pancake-shaped layers and cap-shaped nanostructures exhibited similar mechanical contrasts in the corresponding adhesion, stiffness (DMT modulus), and deformation maps, which were considerably different from those of the ordered base layers. Several mechanical properties of INBs and micropancakes at the HOPG-water interface prepared using the standard solvent exchange method (*2*) were also measured with the PF mode. The similar shapes, dimensions, and mechanical properties indicated that the cap-shaped nanostructures (thick pancake-shaped layers) were probably the INBs (micropancakes).

When nitrogen-supersaturated water was deposited onto a clean HOPG sample, cap-shaped nanostructures and patches of monolayer ordered structures were observed at the HOPG-water interface. A typical example is shown in Fig. S5. A more detailed characterization of the interfacial structures using the PF mode is shown in Fig. 3. The ordered structure was oriented along one of three equivalent zig-zag directions of the HOPG substrate, the lattice of which was imaged using the contact mode (Fig. S6). The structure exhibited the same appearance as that reported at the HOPG-water interface using pre-degassed water in a nitrogen or air environment (*36-39*). Figure 3A is a height image acquired at a peak force of 150 pN. No specific pattern was observed inside the cap-shaped structure in the corresponding adhesion map (Fig. 3B), suggesting a disordered structure.

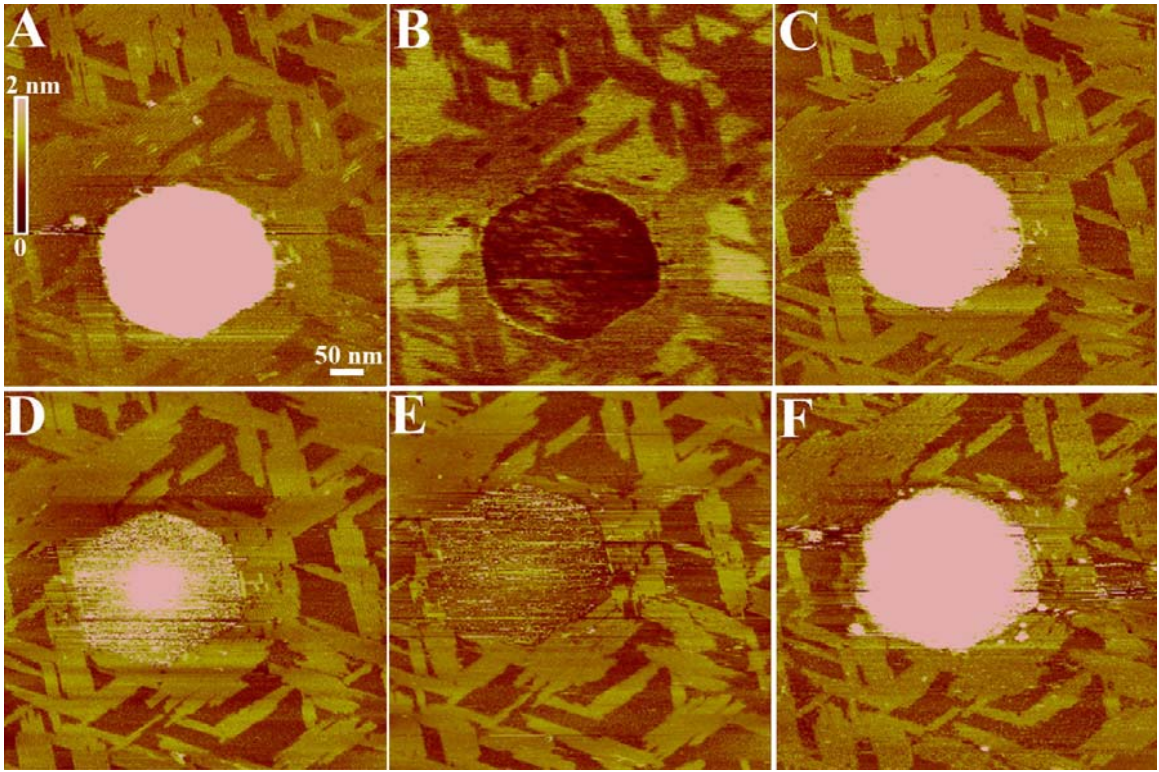

**Fig. 3. Images of an HOPG-water interface acquired using the PF mode at approximately 4 hours after nitrogen-supersaturated water was deposited**. (A), (C), (D), (E), and (F) are height images. The peak force used to acquire (A) and (B) was 150 pN. Peak forces of 750 pN (C), 1500 pN (D), and 2250 pN (E) were applied and, finally, the force was reduced to 150 pN (F). The apparent heights at the center of the cap-shaped structure in (A), (C), (D), (E) and (F) were 18.5 nm, 10.2 nm, 2.7 nm, 0.2-0.4 nm, and 12.6 nm, respectively. The left edge of the 3D nanostructure in (F) exhibited several changes after applying a peak force of 2250 pN. The real height of the cap-shaped nanostructure after the PF measurements was 20.3 nm, which was determined from the force curve measurements conducted at the center of the structure (see Fig. S7). The domains of the row-like structure changed slowly with time. Furthermore, several domains exhibited damage at high peak forces (e.g., domains near the left edge of the cap-shaped structure). (B) is the corresponding adhesion map acquired along with (A). Both the cap-shaped nanostructure and the monolayer ordered structure appeared darker than the bare HOPG substrate. In (F), an increasing number of bright particles with heights ranging from 0.8 nm to 3 nm appeared after scanning with a high peak force.

As shown in Figs. 3A-3E, the apparent height of the cap-shaped nanostructure decreased as the peak force was increased. When the peak force reached 2250 pN, the apparent height decreased to nearly zero (Fig. 3E) and no underlying structure was observed, indicating that the bare HOPG substrate was under the cap-shaped nanostructure. The cap-shaped nanostructure was surrounded by domains of the row-like pattern, forming the hexagonal boundaries of the structure. When the peak force was reduced from 2250 pN to 150 pN, the cap-shaped structure reappeared (Fig. 3F). However, the apparent height decreased from the 18.5 nm shown in Fig. 3A to 12.6 nm. This probably occurred because scanning at the maximal peak force of 2250 pN strongly perturbed the structure, causing dissolution or removal of gas molecules inside the cap-shaped nanostructure. The increasing number of small bright particles that appeared on the surrounding row-like structure in Fig. 3F might have been engendered by the gas inside the cap-shaped nanostructure and the destruction of areas of the monolayer ordered structure. These particles frequently changed sites and disappeared after one to several scans. This observation is similar to our recent FM-AFM observations at the HOPG-water interface in a nitrogen environment (*39*).

## *Schematics of interfacial structures and the role of interfacial water*

Based on the aforementioned observations and many similar experiments conducted at the HOPG-water interface, schematics for the interfacial structures of nitrogen and oxygen are illustrated in Fig. 4 (**Supplementary text S3**). This work showed that the thick pancake-shaped layers are primarily dense disordered layers of oxygen adsorbates on top of the ordered base layers of oxygen. This discovery explains a previous report that micropancakes form on crystalline hydrophobic substrates (*18*), because the epitaxial base layers can form only on crystalline substrates. We speculate that the structures in quasi-2D disordered layers have a certain degree of short-range order in the lateral directions that is attributable to interaction with the underlying ordered base layers; however, structural fluctuations occur faster than our AFM imaging rate (~1 ms per pixel or ~0.5 s per line). Researchers have recently suggested that micropancakes are 2D dense gas adsorbates (*9, 10*). Our observations provide evidence supporting this suggestion, but the underlying ordered base layers were not expected.

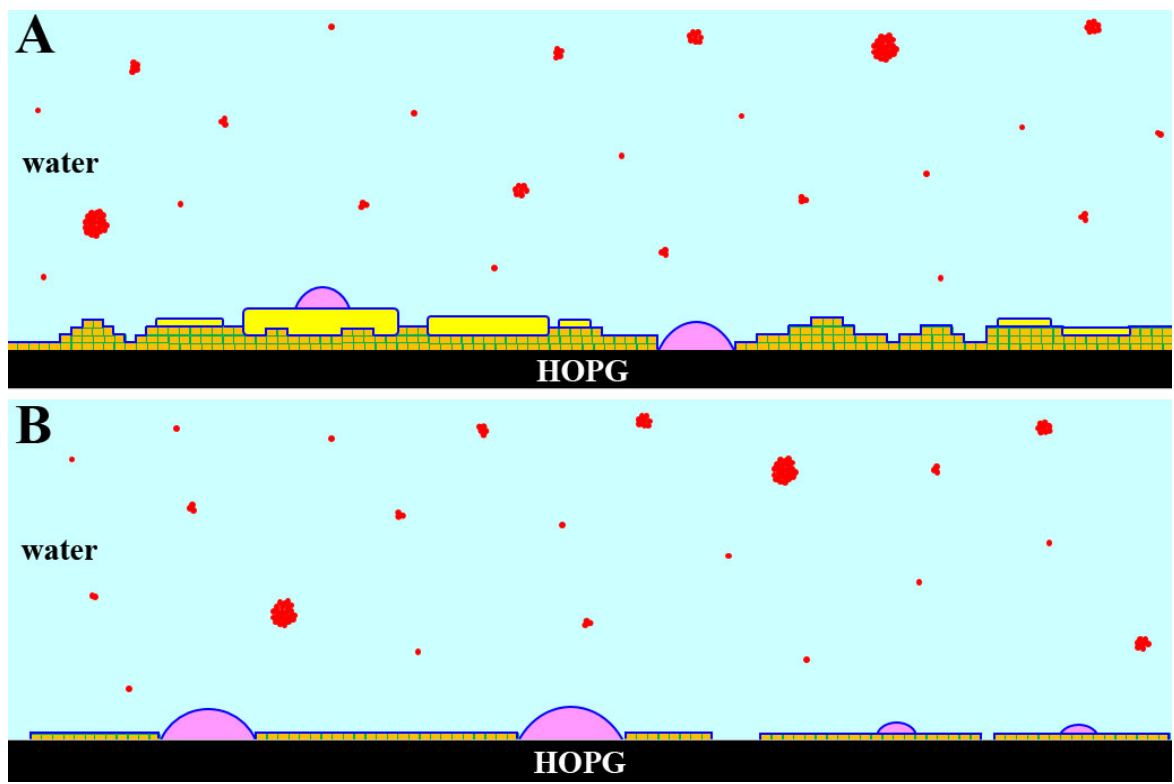

**Fig. 4. Schematic of the structures at the HOPG-water interface when water is supersaturated with pure oxygen (A) and pure nitrogen (B).** For clarity, some structures may not be represented in the correct scale. Orange squares represent ordered structures; cap-shaped nanostructures and small nitrogen and oxygen clusters at the interface are shown in purple; 2D disordered layers are shown in yellow; and gas molecules dissolved in water are shown in red. Each red sphere represents an individual gas molecule and some molecules can aggregate into clusters in water. The interfacial water layers that stabilize these 2D and 3D nitrogen and oxygen structures are shown in deep blue.

Through strong perturbation using the AFM tip, a micropancake can be transformed into an INB with no significant change in the total volume (*17*), indicating that the material density of the two structures is similar. This finding is consistent with those of the current study, which revealed that the various mechanical properties of these two structures measured using the PF mode and force curves are similar. Thus, the cap-shaped nanostructures are probably 3D disordered structures containing close-packed gas molecules. They can be considered 3D liquid-like nanodroplets comprising gases rather than bubbles in the low-density vapor phase (**Supplementary text S4**).

The observation of INBs on micropancakes indicates that the chemical potentials of gas molecules inside the two structures were similar (*16*). INBs can be moved by using an AFM tip on the top plateau of a micropancake (*15*) and two INBs on the same micropancake may coalescence into a larger INB (*16*). A thin water layer was suggested to separate these two structures (*15*). The current study revealed no notable gas transfer among the cap-shaped nanostructures, the pancake-shaped layers, and the ordered adlayers, as well as between these interfacial structures and the surrounding water over a timespan of several minutes. The interfacial water layers probably stabilized the interfacial structures of gases by reducing the interfacial energy. They also act as a barrier to hinder gas transfer among the interfacial structures and between the structures and surrounding bulk water (**Supplementary text S5**). This explains several experimental observations presented in this paper and other AFM studies. Interfacial water layers require further experimental and theoretical investigations.

## *Thermodynamic stability of the interfacial structures of gas molecules*

A recent TEM study reported that hydrogen INBs can be generated in an aqueous solution by using electron radiolysis (*13*). The INBs were estimated to have a high gas density at room temperature. If INBs are in a "dense gas" phase, high pressure is expected, and induces strong water flow from INBs. However, tracer particle tracking measurements indicated no water flow around the INBs (*11*). In addition, INBs have a planar interface with micropancakes and thus no pressure drop occurs between INBs and micropancakes. Micropancakes are 2D interfacial structures of which the internal pressure should be equal to the pressure in the surrounding water, approximately 1 atm. We thus conclude that the pressure inside INBs is not much higher than 1 atm.

Numerous experiments suggest that INBs at hydrophobic-water interfaces are probably thermodynamically stable under ambient conditions. For example, INBs can be formed in water with air near saturation level (*10*) and most INBs are stable for hours or days (*1-12*). These facts indicate that the chemical potentials of gases in INBs are equal to or lower than those in ambient air. This precludes the possibility that INBs are hyperbaric gaseous bubbles because gases inside pressurized bubbles have higher chemical potentials than gases in ambient air do (**Supplementary text S6**).

Our previous AFM observations have indicated that dissolved nitrogen molecules can be adsorbed at the HOPG-water interface and form a monolayer row-like structure even in under-saturated water (*36-39*), suggesting that HOPG, a hydrophobic solid substrate, in water may provide sites of low chemical potential at which dissolved gas molecules can be adsorbed. This is probably because the attractive VDW interactions between gas molecules and the hydrophobic substrate, as well as the interactions between adsorbed gas structures and the interfacial water, are responsible for reducing the chemical potentials. This mechanism also explains the formation and high stability of the various interfacial structures of gases at the hydrophobic-water interfaces.

## *Relevance to the nanoprecipitates of inert gases in solids*

Inert gas atoms exhibit low solubility in most solids and form nanoprecipitate when implanted into solids; this behaviour is analogous to the behavior of nonpolar gas molecules in water. The most unexpected discovery was the extremely high atomic densities (equivalent to those in solid or liquid states) of inert gas atoms within the small cavities of the host solids at room temperature (*19-35*). In particular, solid phase precipitation of the heavy inert gases (Ar, Xe, Kr) in metals (*24-33*) and crystalline Si have been reported (*35*). The solid nanoprecipitates exhibited a structure epitaxial with the host matrix. The most commonly studied system is that of Xe in Al thin films of approximately 50 nm. Xe can be condensed into a close-packed solid state when confined in Al cavities with diameters smaller than 6 nm and into a liquid state when confined in a cavity larger than 10 nm (*30-32*). High pressure (several to hundreds of thousands of atmospheres) was expected in the nanoprecipitates based on the phase diagrams of bulk gases. The Young-Laplace equation was applied to explain the solid and liquid phases of precipitates of different sizes, because a higher Laplace pressure is expected inside a smaller spherical cavity (**Supplementary text S7**).

This study showed that ordered epitaxial adlayers of oxygen and nitrogen, solid-gas phases, can form at an HOPG-water interface under ambient conditions. HOPG is atomically flat and, thus, no high pressure is expected inside the ordered gas adlayers. This proves that compressive high pressure is not essential for the formation of the solid-gas phase. A TEM study reported that semicrystalline Xe precipitates exhibited a solid phase in the first three layers adjacent to the crystalline Al substrate and a liquid phase in their center (*32*). This implies that the solid structure of Xe atoms is induced by the crystalline substrate; this process is analogous to the formation of the first one to four ordered adlayers of nitrogen and oxygen at the HOPG-water interface. Strong interactions between gas and a solid surface relative to the gas-gas interactions are a key factor in the ordering of gas in our system as well as in inert-gas precipitates inside solids.

The VDW interaction energy between gas molecules can be estimated based on the latent heat of sublimation (*41*) or the viscosity coefficient (*42*). Gas-solid interactions have been a key research topic in surface science for decades (*43-45*). One rarely-known fact is that gas-solid interactions are often considerably stronger than gas-gas interactions (by one order of magnitude), as shown in Table 1. For example, the energy values of $N_2$–$N_2$, $N_2$–HOPG, $O_2$–$O_2$, $O_2$–HOPG, Xe–Xe, and Xe–metal interactions are approximately 9 meV, 100 meV, 10 meV, 100 meV, 22 meV, and 200–400 meV, respectively. The strong interfacial interaction between gas molecules and solid substrates is conducive to condensation of the gases confined in solids or at a solid–liquid interface. In addition, the crystalline substrate provides a periodic corrugation of holding potentials for the adsorbates, which may drive the solid ordering of the gas molecules. This interface-induced ordering weakens when the molecules are located farther from the interface. Thus, the ordered adlayers persist for only a few atomic or molecular layers, explaining our observation of the first one to four ordered adlayers at the HOPG−water interface and previous observations of the solid–gas precipitates inside the small cavities of the solids (**Supplementary text S8**).

| Gas-solid | Potential Well Depth (meV) (*44*) | Gas-Gas | 6-12 Lennard Jones Potential Well Depth (meV) (*42*) | Dispersion Interaction (meV) (*41*) |
|---|---|---|---|---|
| $N_2$-Graphite | 104±3 | $N_2$-$N_2$ | 7.89 | 9.1 |
| $O_2$-Graphite | 90.4~101.7 | $O_2$-$O_2$ | 9.75 | 10.5 |
| Ar-Graphite | 96±2 | Ar-Ar | 10.70 | 11.1 |
| Ar-Ag(111) | 72±7 | | | |
| Xe-Graphite | 162±4 | Xe-Xe | 19.75 | 21.5 |
| Xe-Cu(111) | 172~194 | | | |
| Xe-Pd(111) | 356 | | | |
| Xe-Pt(111) | 269~304 | | | |
| Xe-W(110) | 180±6 | | | |
| Xe-Al(111) | 176 (*45*) | | | |
| He-Graphite | 16.2~17.0 | He-He | 0.88 | |
| He-Ag(111) | 6~7 | | | |
| He-Cu(110) | 6.27±0.8 | | | |
| He-W(110) | 3.5 | | | |
| He-Pt(110) | 8.5±0.2 | | | |

**Table 1. Gas-solid and gas-gas interactions**

## *Discussions*

Hwang et al. proposed that the formation of a gas agglomerate (a gas cluster) in water or at a solid-water interface is energetically more favorable than the formation of a gaseous bubble when the number of confined gas molecules, *N*, is lower than a critical size $N_c$ (*36, 38*). This concept can be understood qualitatively. For a small structure, the ratio of interfacial (surface) area to volume increases as the size decreases, and the contribution of the interfacial (surface) energy to the total free energy becomes more dominant. Therefore, the thermodynamic properties of gas molecules confined in a medium (such as a solid, water, a solid-water interface, or other materials in a condensed state) of a sufficiently small space can be considerably different from those of bulk gases. Interfacial energy is generally positive and roughly proportional to the interfacial area, which favors a condensed state rather than a gaseous state (*36, 38*). In addition, the energy per unit area for cluster-water (cluster-solid) interfaces should be substantially lower than that for vapor-water (vapor-solid) interfaces because of the additional attractive VDW interactions for the former interfaces. The critical size, $N_c$, increases as the interfacial interaction between gas molecules and the medium becomes stronger. This concept can be generalized to a precipitate or a material embedded inside another material or located at interfaces of other materials. This indicates that the thermodynamic properties of an inclusion may be dominated by the interaction energy at the interface when its size is sufficiently small.

INBs are 3D liquid-like gas structures, which may have distinct physical properties. The mutual interactions among gas molecules are merely weak VDW forces, and thus INBs may shear easily and behave as a low-viscosity liquid. This can reduce the friction for motion of a solid above INBs and drag for water flow above. The presence of a low-viscosity layer at the solid-water interface was suggested to explain the boundary slip and drag reduction of water flow (*46*), but the physical origin of the low-viscosity layer remains unclear. The VDW liquid in INBs may provide a clue that can be used in determining the mechanism. The findings and explanations presented in this paper may clarify numerous puzzles regarding gases confined in a small space and have implications regarding various research fields. For example, hydrophobic surfaces can effectively catch gases dissolved in water; this property may have implication regarding the breathing of marine species and other phenomena. In addition, gases dissolved in aqueous solutions may affect the self-assembly and functions of biological molecules. Furthermore, interface-induced ordering provides new possibilities for high-density gas storage.

**Acknowledgments:**

We thank the National Science Council of the R.O.C. (NSC99-2112-M-001-029-MY3 and NSC102-2112-M-001-024-MY3) and Academia Sinica for supporting this study.

# Supplementary Materials for:

## Interface-induced ordering of gas molecules confined in a small space

Yi-Hsien Lu, Chih-Wen Yang, Chung-Kai Fang, Hsien-Chen Ko, Ing-Shouh Hwang*

**Materials and Methods:**

**Atomic force microscopy.** The images in this paper were acquired using a Bruker AXS Multimode NanoScope V equipped with a commercial fluid cell tip holder. A schematic of the setup is shown in Fig. S8. The oscillation of the cantilever was driven using a dynamic-modulation system composed of a lock-in unit/phase-lock-loop (PLL) unit (Nanonis OC4 Station from SPECS) and a signal access module (Bruker AXS). For the frequency-modulation (FM) mode, the PLL unit was employed to track the resonance frequency of the vibrating cantilever. The resonance frequency shift ($\Delta f$) was used as the feedback input signal of a proportional-integral controller to obtain topographic images. The driving amplitude, which was controlled using the PLL to maintain a constant cantilever oscillation amplitude during scanning, could be measured and recorded simultaneously. This value was typically considered the energy dissipation signal (*47*). A dissipation map was obtained along with the topographic images. The FM detection scheme was used here because it has exhibited superior force sensitivity compared with the conventional tapping mode (*36-38, 48-50*). The sample displacement was driven using a piezoscanner under the sample stage. Si cantilevers (OMCL-AC240TS from Olympus) with a spring constant 0.7-3.8 N/m were used. The nominal tip radius was approximately 10 nm. The resonance frequency of the cantilever was approximately 32 kHz in water. For operation using the FM mode, the resonance amplitude was 3.6 nm and the resonance frequency shift was set at +10 to +15 Hz.

**PeakForce (PF) mode.** Regarding the PeakForce mode, the sample was oscillated in a vertical direction with an amplitude of tens to hundreds of nanometers and at a frequency of 2 kHz (Fig. S8B). The vertical piezo movement results in cycles of approaching and retracting traces in which the tip makes intermittent contact with the sample surface. A force-distance curve was acquired in each cycle. Topography information was obtained from the height correction performed by the feedback loop to keep a constant "peak" of force, whereas the slope of the force curve at the contact region determined the stiffness of the sample at each pixel. Other information on the surface mechanics, such as adhesion, deformation, or dissipation, was obtained using the measured cycle of the approach-retraction force curve, as shown in Fig. S8B (*51, 52*). In the operation, the scan rate was approximately 1 Hz and the oscillation amplitude was 15-25 nm.

**Materials and sample preparation.** The highly ordered pyrolytic graphite (HOPG) samples (lateral sizes of 12 mm × 12 mm, ZYB; Momentive) were cleaved prior to each AFM experiment. All water was purified using a Milli-Q system (Millipore Corp., Boston) with a resistivity of 18.2 MΩ·cm. In preparation of water supersaturated with nitrogen (oxygen), a beaker containing DI water (30-50 mm high) was placed in a pressure tank (TNKB1-3; Misumi) which was pressurized to 2.4-3.7 atm with pure nitrogen (oxygen). An air filter (MSAF8A-0.01, Misumi) with a filtration level of 0.01 μm was used between the high pressure gas cylinder and the stainless-steel pressure tank. The water was stored for several days before being opened for the AFM experiments. All of the experiments were conducted at room temperature.

**Supplementary Text:**

**S1: Young-Laplace equation and puzzles about INBs.**

The Young-Laplace equation implies a high internal pressure for gas inside a bubble of a small radius, which should drive gas diffusion across the interface and cause the bubbles to disappear in a very short time. The lifetime of a gas bubble thus decreases with decreasing bubble size. According to this equation, a gas bubble in water having a radius of 100 nm would have a pressure ~15 atm and should disappear in less than 1 ms (*14*). Hence it is very surprising that the INBs observed by AFM are stable for hours or days (*1-12*). Several mechanisms have been proposed (*8, 9, 12, 40*). In these models, INBs are not thermodynamically stable and can have an extended lifetime only under certain conditions. They can explain neither the reported superstability of INBs (*53*), nor the experimental observations that most INBs exist for more than four days and even grow in size with time (*7, 12*).

**S2: Micropancakes.**

Micropancakes were first proposed to be in a novel gaseous state (*15-18*). It remains mysterious why their interface with water is not spherical or semi-spherical, as the water-vapor interface has a very high free energy per unit area. Micropancakes were also proposed by some researchers to comprise dense gas adsorbates (*9,10*), but there was lack of experimental evidence. In particular, INBs can sit stably on micropancakes, which implies a similar chemical potential for gases inside these two structures. It is difficult to explain why a gaseous nanobubble with a high Laplace pressure can sit stably on a dense gas adsorbate layer.

**S3: Models for interfacial structures of nitrogen/oxygen at the HOPG-water interface.**

Figure 4A illustrates a schematic when water is super-saturated with oxygen. The HOPG substrate has been covered by one to four molecular layers of ordered structures. On top of them, there might be disordered layers of varying thicknesses, from one molecular layer to several nm. The thick disordered layers are the pancake-shaped structures in AFM images, or micropancakes. The disordered layers are mainly composed of adsorbed $O_2$ molecules that are constantly undergoing structural transformations, disordering. Cap-shaped nanostructures can form on top of the thick disordered layer. Figure 4B illustrates a schematic when water is super-saturated with nitrogen. The interface is mainly composed of domains of a molecular layer of an ordered structure. The cap-shaped nanostructures can form on bare HOPG substrate at regions without the ordered layer, but small gas clusters can be seen occasionally on the ordered monolayer. These different characteristics shown in Figs. 4A and 4B provide evidence that the observed structures do not come from contaminants in DI water.

As illustrated in Fig. 4, dissolved gases are in the form of monomers or clusters of various sizes in water super-saturated with nitrogen or oxygen. This is based on the comparison of the current study using nitrogen-supersaturated water with another recent study of ours using pre-degassed water in a nitrogen environment (*39*). In the latter study, where dissolved $N_2$ molecules can be expected to be mainly in the well-dispersed monomer form in water, the same row-like pattern remains but no cap-shaped nanostructures are present on bare HOPG even when the entire interface has been covered by the ordered structure. We thus conclude that the ordered structures are formed through self-assembly of individual nitrogen molecules (perhaps also including clusters containing only a small number of molecules). In the present work, where nitrogen-supersaturated water is used, the cap-shaped nanostructures appear on the bare HOPG substrate when the ordered layer covers only about 50% of the interface. This strongly suggests that a portion of dissolved nitrogen molecules probably aggregate into large clusters in super-saturated water. Adsorption of a large nitrogen cluster onto the bare HOPG substrate may lead to nucleation of a cap-shaped nanostructure. Further adsorption of nitrogen, either in the dispersed form or in the cluster form, leads to its growth. We note that previous theoretical studies have indicated that small non-polar molecules can aggregate into clusters of nanometer sizes in water (*54, 55*).

For many experiments conducted under water super-saturated with oxygen, occasionally we observed cap-shaped nanostructures on bare HOPG substrates. In that situation, no underlying ordered base layers were observed at high peak forces, similar to the case shown in Fig. 3E. Thus adsorption of cap-shaped oxygen nanostructures on bare HOPG is also illustrated in Fig. 4A.

**S4: INBs might be nanodroplets of gases.**

An earlier experimental report also supports a condensed state in the INBs. Nanodroplets of decane were prepared at a hydrophobic-water interface using the solvent exchange procedure (*56*), which is also a widely used method to prepare INBs at solid-water interfaces (*2*). Their shape and dimensions observed by AFM are very similar to the INBs containing air or gas. The nanodroplets of decane are very flat and their contact angle is very similar to that of macroscopic droplets of decane at the same interface (*56*). In comparison, the INBs are also very flat with a small contact angle (measured from the gas side) but the macroscopic gas bubbles at the solid-water interfaces have much larger contact angles. This suggests that INBs might be a liquid-like nanodroplets containing gas and that there may be a phase transition between INBs and macroscopic gas bubbles.

**S5: Interfacial water stabilizes the interfacial structures of gases.**

Interfacial water may play a crucial role in stabilizing the interfacial structures of gases and forming a barrier separating different interfacial structures or separating the interfacial structures from the bulk water. Thus the gas transfer rates between different interfacial structures and between the interfacial structures and the surrounding water are very small. Interfacial water may also confine the oxygen molecules in micropancakes laterally, which can explain why they exhibit a clear boundary in the lateral directions (*17*). Our observations have indicated that the ordered base layers nucleate and grow on the time scale of minutes or longer. When the ordered row-like structures are destroyed using the tip, the restructuring of new ordered structures requires a similar time scale. Since we cannot see individual nitrogen/oxygen molecules at the HOPG-water interface, these adsorbed monomers probably diffuse rapidly at the interface at a rate far above our AFM imaging speed. This suggests that there are significant energetical and/or entropical barriers for the arrangement of nitrogen/oxygen molecules into the ordered row-like structures. Considering the large row separation of approximately 4 nm relative to the size of single molecules, 0.2-0.3 nm, the nucleation and growth of a domain of a row-like structure would definitely involve rearrangement of a large number of nitrogen or oxygen molecules. If interfacial water molecules are also involved in the rearrangement, a significant energetical and/or entropical barrier for the assembly of the ordered structures can be expected.

It was recently reported that INBs shrank slowly in size in partially degassed water and grow slowly in gas-super-saturated water on the time scale of many hours (*12*). This clearly indicates a barrier to hinder the gas transfer between INBs and surrounding water, even though the chemical potentials of gases in the surrounding water were changed. The interfacial water may be responsible for this barrier and the surprisingly high stability of INBs.

**S6: Chemical potentials of gases in interfacial structures.**

The chemical potential of a gas species *i* in a vapor (gas) phase can be written as

$$\mu_i^g = \mu_i^{SATP} + RT \ln(\lambda_i p_i) \qquad \text{(A)}$$

where $p_i$ is the partial pressure of gas species *i* in bar, $\lambda_i$ is the fugacity coefficient of gas *i*, $\mu_i^{SATP}$ is the chemical potential of gas *i* in air (such as nitrogen and oxygen) at the standard ambient temperature and pressure (SATP, 25°C and 1 bar), *T* is the temperature, and *R* is the gas constant. This equation indicates that gases of a higher pressure are thermodynamically less favorable. Under thermodynamic equilibrium at SATP, the chemical potential of gas i dissolved in water is equal to $\mu_i^{SATP}$ and the gas concentration reaches the saturation concentration, $c_i^0$. The high stability of INBs indicates that the chemical potential for gas i inside INBs should be equal to or smaller than that in the gas-saturated water, i.e. $\mu_i^{INB} \leq \mu_i^{SATP}$. This precludes the possibility that INBs are hyperbaric gaseous bubbles because, according to Eq. (A), gases inside pressurized bubbles would have chemical potentials higher than $\mu_i^{SATP}$ and tend to dissolve into the surrounding water.

The chemical potential of gas *i* dissolved in water can be written as

$$\mu_i^{sol} = \mu_i^{SATP} + RT \ln(a_i) = \mu_i^{SATP} + RT \ln(\gamma_i c_i / c_i^0), \qquad \text{(B)}$$

where $a_i$ is the activity of the species in solution, $\gamma_i$ is the activity coefficient, and $c_i$, $c_i^0$ are the concentration of gas *i* in water and the saturation concentration at SATP, respectively. That means the chemical potential of gas *i* dissolved in water will be reduced at a lower concentration. In partially-degassed water, no INBs were observed and only domains of a monolayer row-like structure were present at the HOPG-water interface (*36-39*). This indicates that the chemical potential of nitrogen molecules in the row-like structure should be smaller than that in air and in INBs.

**S7: Cause for the condensation of inert gases inside small cavities of solids.**

A key issue is whether high pressures are the cause for the condensation of inert gases inside the small cavities at room temperature. In general, highly pressurized gases have very high chemical potentials and are thermodynamically less favourable. A TEM study showed that the total cavity volume was conserved during coalescence of two crystalline Xe nanoprecipates in Al (*29*), indicating that the Young-Laplace equation based on spherical cavities cannot provide accurate estimation of the pressure inside the precipitates. In addition, through experimental determination of lattice spacing or gas density at a certain temperature, pressure could be calculated through an appropriate equation of state of bulk gases, which describes the relation of pressure-atomic density-temperature. In many cases, these pressure values deviate significantly from the Laplace pressures (*33*). For example, spherical He precipitates of 2 nm can form a superlattice in metals with atomic densities significantly larger than that estimated from the Young-Laplace equation (*22, 23, 33, 34*). This also indicates that the Young-Laplace equation is not appropriate in estimating the pressures inside the spherical He nanoprecipitates. Furthermore, experimental observations have indicated that many cavities are faceted. There should be no pressure difference across these planar facets. Moreover, He nanoprecipitates can form a planar morphology, platelets, inside metals with the atomic density equivalent to that in liquid or solid helium (*21, 33, 34*). Again there should be no pressure difference across the planar interfaces. Moreover, these precipitates exhibit good stability upon annealing to elevated temperatures when they can move, change shapes, or undergo coalescence (*21, 33, 34*).

**S8: Solid Xe precipitates inside small cavities of Al.**

Entropy may also contribute to the size-dependent order-disorder transition. Because of the very limited possible configurations for the gas arrangement inside a small cavity filled with close-packed gas molecules, the entropy gain caused by disordering is smaller as the cavity size decreases. Thus disordering is less favorable for a smaller cavity and the order-disorder transition temperature can be expected to be higher. Along with the interface induced ordering, the solid Xe in Al cavities smaller than ~6 nm and liquid Xe in larger cavities can be explained.

**S9: Comparison with $N_2$/HOPG and $O_2$/HOPG systems in vacuum.**

For the last several decades, many systems of gas adsorption on crystalline solids have been studied with surface science techniques in vacuum, including $N_2$/HOPG and $O_2$/HOPG systems (*57-59*). Several ordered structures for $N_2$/HOPG and $O_2$/HOPG systems have been observed at different partial pressures of nitrogen and oxygen at temperatures below room temperature. Those structures are very different from the row-like patterns at HOPG-water interface observed at room temperature. The major difference of the present system compared with the previous systems in vacuum is the additional interface with water. Thus we think the interfacial water, which may form an ordered hydrogen-bonded network, plays a crucial role in stabilizing the row-like patterns, rather than other arrangement of nitrogen/oxygen at the interface.

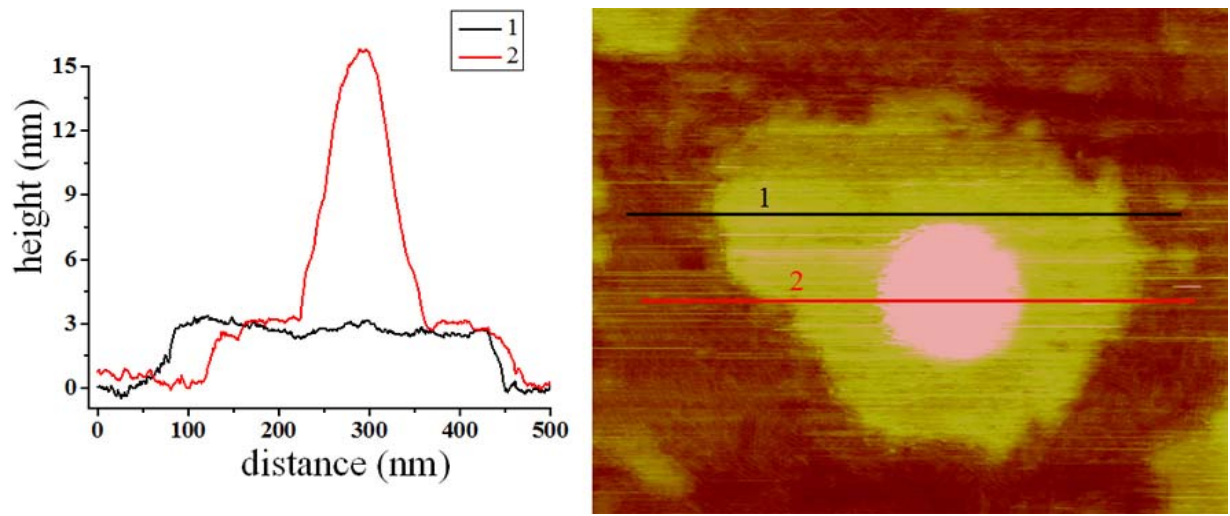

**Fig. S1.** Height profiles across a pancake-shaped structure and across a cap-shaped nanostructure in the topographic image acquired using the FM mode.

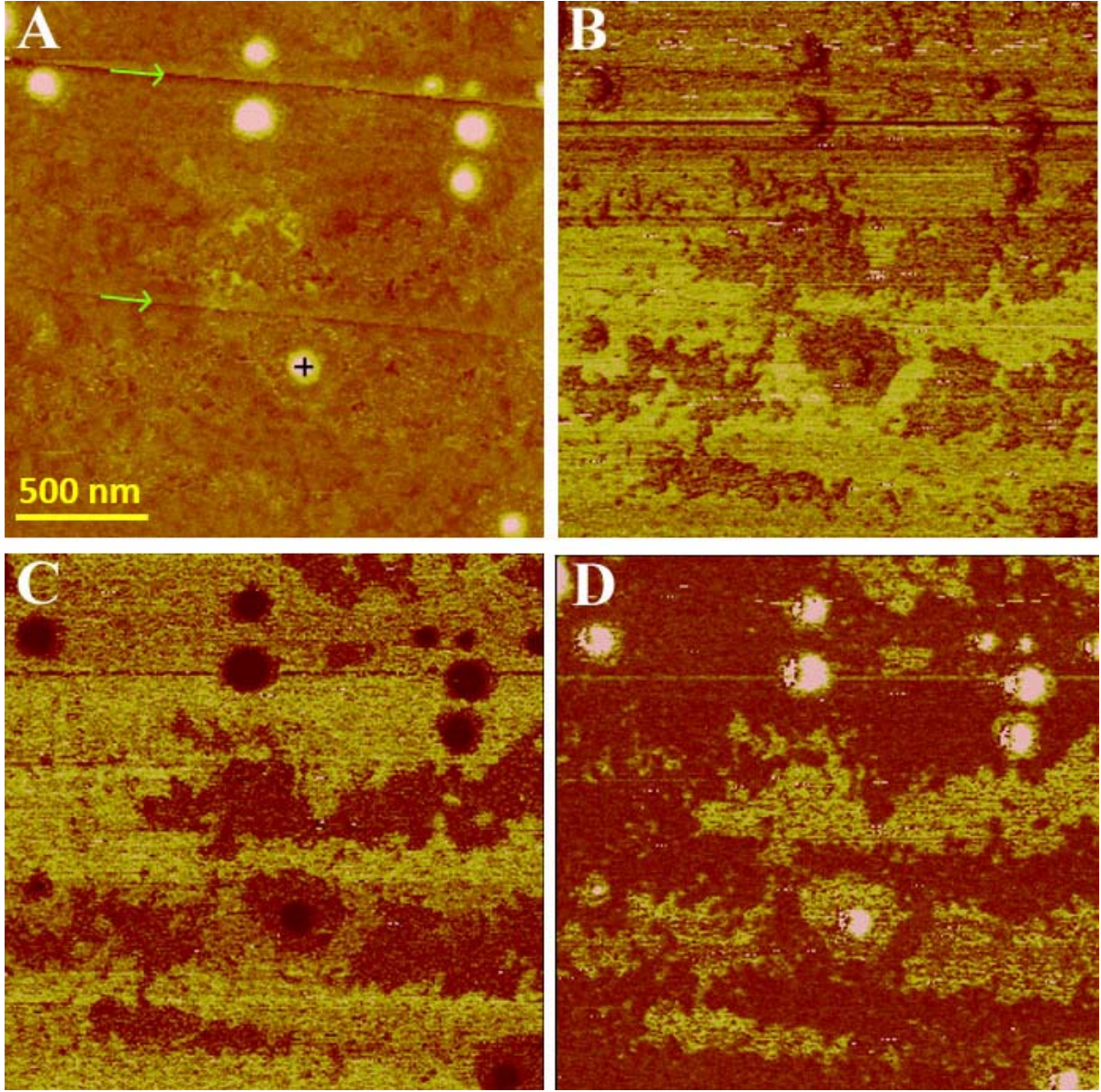

**Fig. S2** PF measurements of a HOPG-water interface at t=340 min. after water supersaturated with oxygen was deposited. The peak force was set to 100 pN, which is near the minimal value that enables stable imaging of the entire area. The oscillation amplitude was 20 nm. (A) Height image acquired at approximately the same location as that shown in Fig. 1A. Two substrate step edges are indicated using green arrows. (B) The corresponding adhesion map. The cap-shaped nanostructures and pancake-shaped layers exhibited lower adhesion than the ordered base layers did. (C) The corresponding stiffness map. The cap-shaped nanostructures appeared slightly darker (softer) than the thick pancake-shaped layers. The base layers were relatively stiff. The lateral sizes of several pancake-shaped layers were substantially greater than the cap-shaped nanostructure on top of the layers, but other layers were only slightly larger. (D) The corresponding deformation map. The cap-shaped nanostructures exhibited the greatest deformation and the thick pancake-shaped layers also exhibited greater deformation than the base layers did.

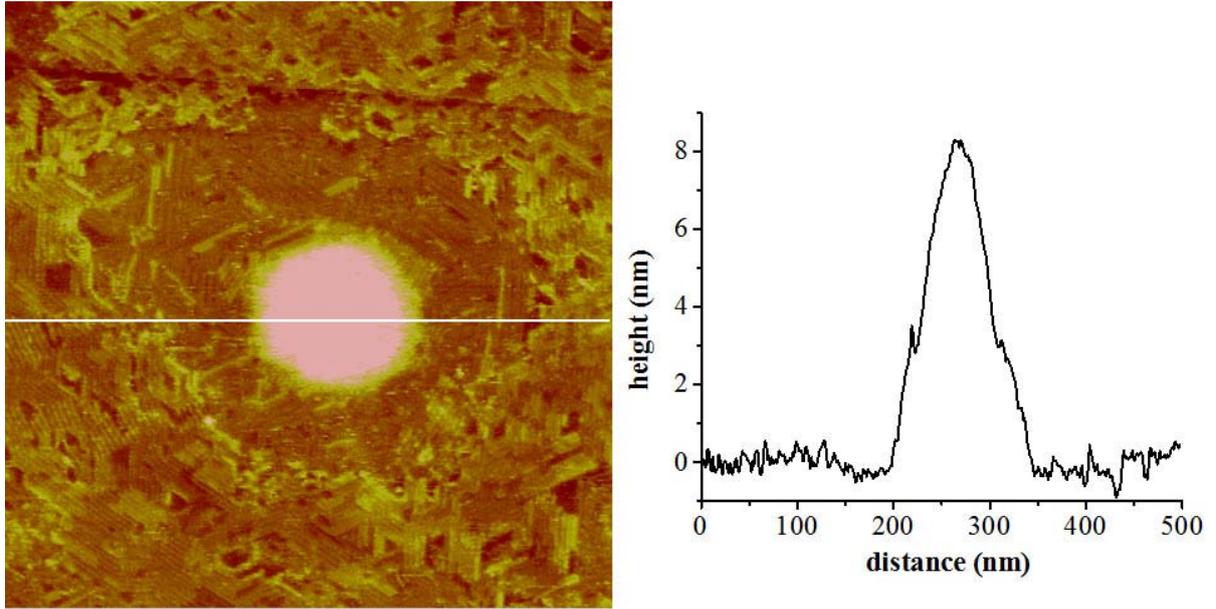

**Fig. S3.** Line profile across a pancake-shaped structure and a cap-shaped nanostructure on the height image shown in Fig. 2A.

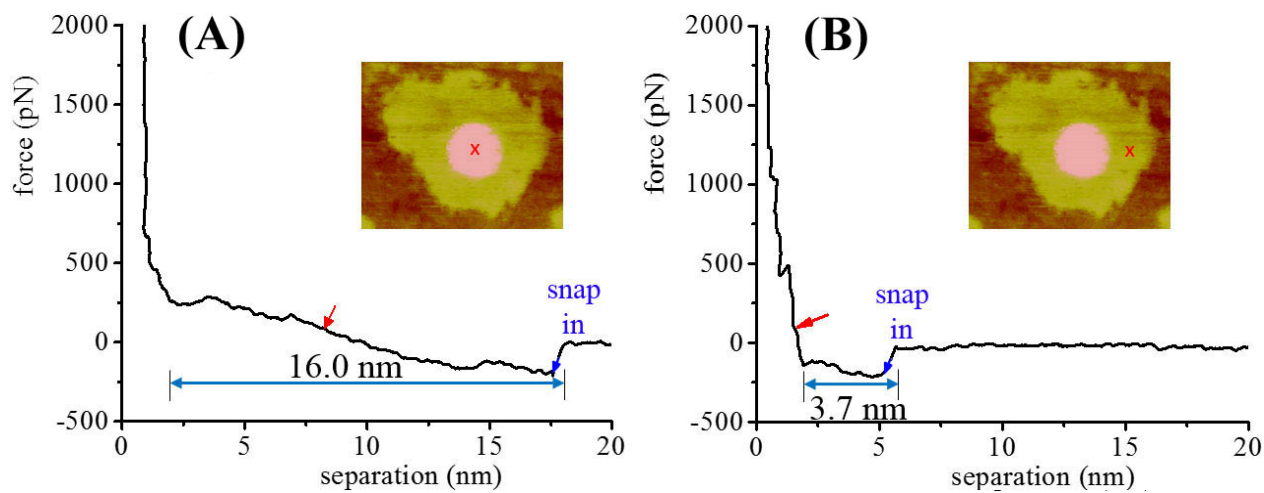

**Fig. S4.** Force vs tip-sample separation curve measured on (A) a cap-shaped nanostructure and (B) a thick pancake-shaped layer when an AFM tip approached the interface. The red cross indicates the position of the force-curve measurements. The ramp velocity was 100 nm/s. Before the tip contacted the interfacial structures, no force was detected. A snap-in occurred when the tip touched the cap-shaped nanostructure or the thick pancake-shaped layer. The soft compliance regions outlined by double-headed arrows enabled us to measure the thickness of the disordered regions in (A) the cap-shaped nanostructure and (B) thick pancake-shaped layer. When the tip reached the stiff base layers, the slopes of the force curves exhibited a sharp increase. The red arrows in (A) and (B) indicate the feedback setpoint of the topographic imaging. The setpoint was evidently located at a certain depth below the surface of the structures. A positive peak force was required for stable AFM imaging of the entire surface; thus, the tip needed to pierce the structures to a certain depth to offset the attractive snap-in force. As indicated by the red arrow in (A), the tip traced a profile that was a certain depth below the surface profile of the cap-shaped structures; thus, the apparent heights obtained using the PF mode were generally smaller than the actual heights. In imaging of the thick pancake-shaped layers, the tip similarly penetrated the thick disordered layer to scan the structures of the underlying stiffer base layers, as indicated by the red arrow in (B). The force-curve measurements in (A) and (B) indicated that the thickness of the thick pancake-shaped layer was 3.7 nm and the height of the cap-shaped nanostructure to was approximately 16 nm. These values are near the measured values, 2.5-3.5 nm and 15.2 nm, respectively, in the topographic images obtained using the FM mode (Fig. S1). This clearly indicated that the FM mode is more accurate in detecting the surface profile of the interfacial structures than the PF mode. By contrast, the PF mode provides clearer imaging of the ordered structures under the disordered structures of gases.

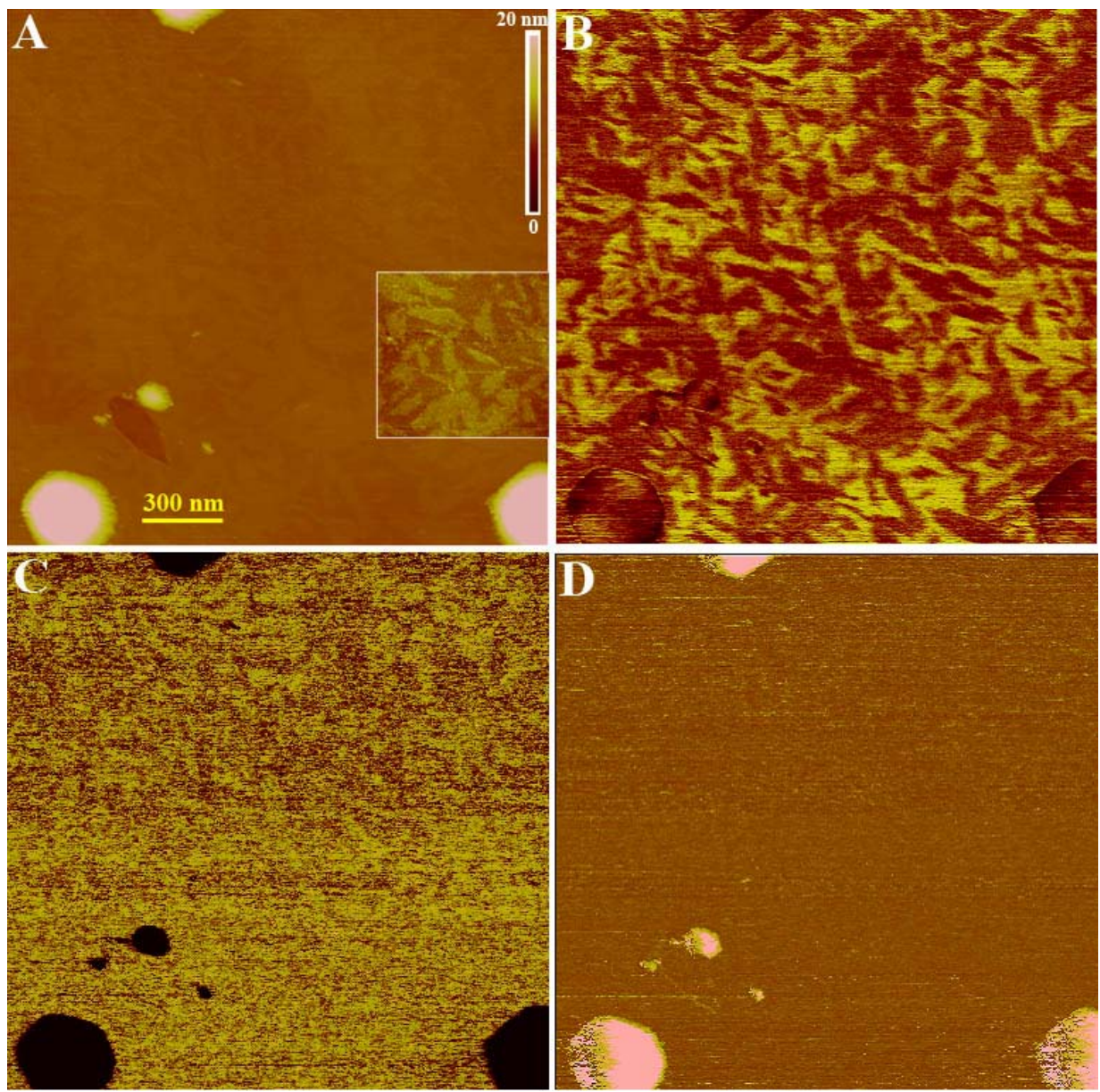

**Fig. S5.** PF maps of a HOPG sample acquired at 115 pN after water supersaturated with nitrogen gas was deposited. (A) Topography map, (B) adhesion map, (C) stiffness (DMT modulus) map, (D) deformation map. The interface was 55% covered by patches of a monolayer ordered structure but cap-shaped nanostructures were distributed sporadically across the interface. Domains of a monolayer row-like structure in the outlined region in (A) were observed clearly because the height contrast was adjusted for this local region. In (B), patches of the monolayer row-like structure and the 3D nanostructures were observed, and adhesion was lower than that of the area of bare HOPG. In (C) and (D), a substantially lower modulus and greater deformation were observed on the cap-shaped nanostructures than on the ordered structures and bare HOPG. The edges of each cap-shaped nanostructure exhibited a hexagonal shape.

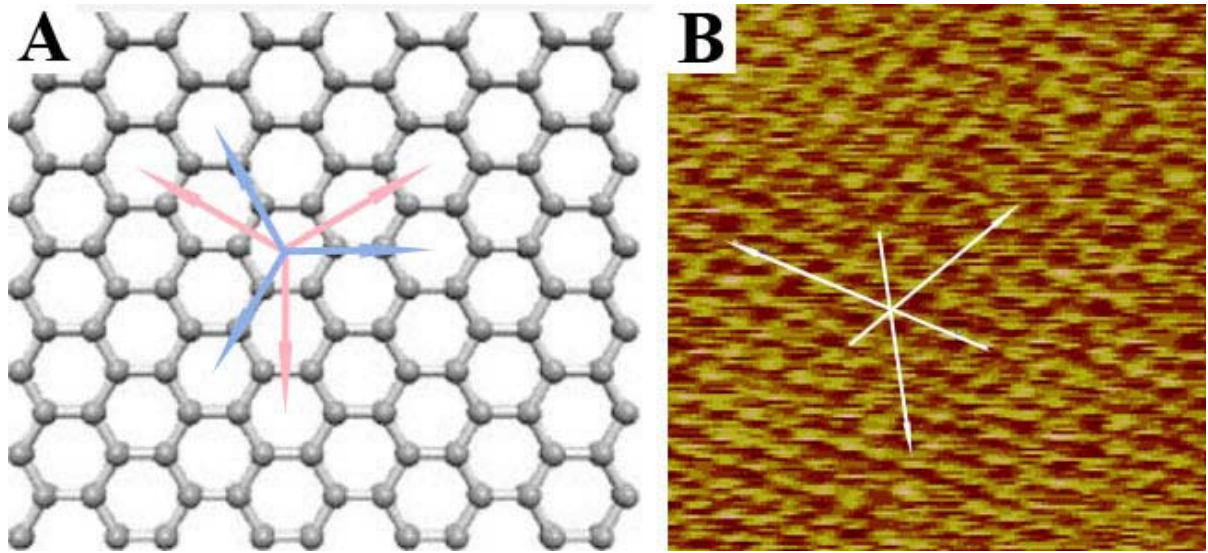

**Fig. S6.** HOPG lattice. (A) Atomic model of the HOPG lattice. Three equivalent zig-zag directions are indicated using pink arrows, and the arm-chair directions are indicated using blue arrows. (B) AFM image of the HOPG lattice in water acquired by scanning at a high loading force and scanning speed by using the contact mode. The three equivalent zig-zag directions are indicated using white arrows. The image was acquired immediately after all PF measurements shown in Fig. 3 and S5 were conducted. The substrate crystal orientation is the same as that shown in those images.

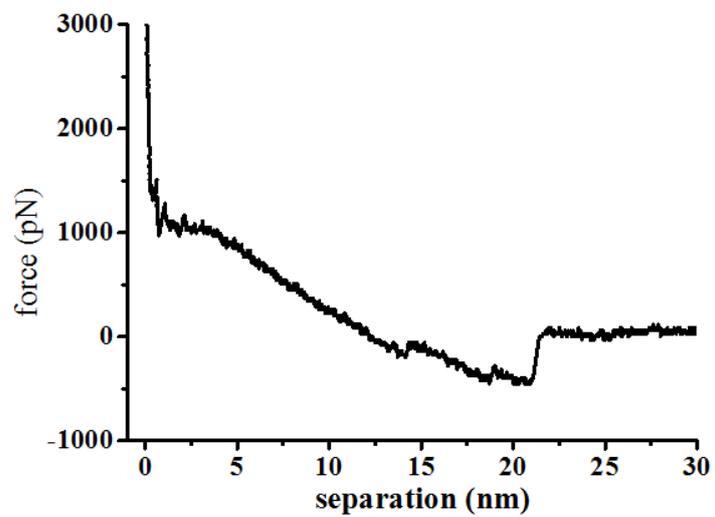

**Fig. S7.** Measurement of force versus tip-sample separation while the tip approached to the center of the cap-shaped nanostructure in Fig. 3.

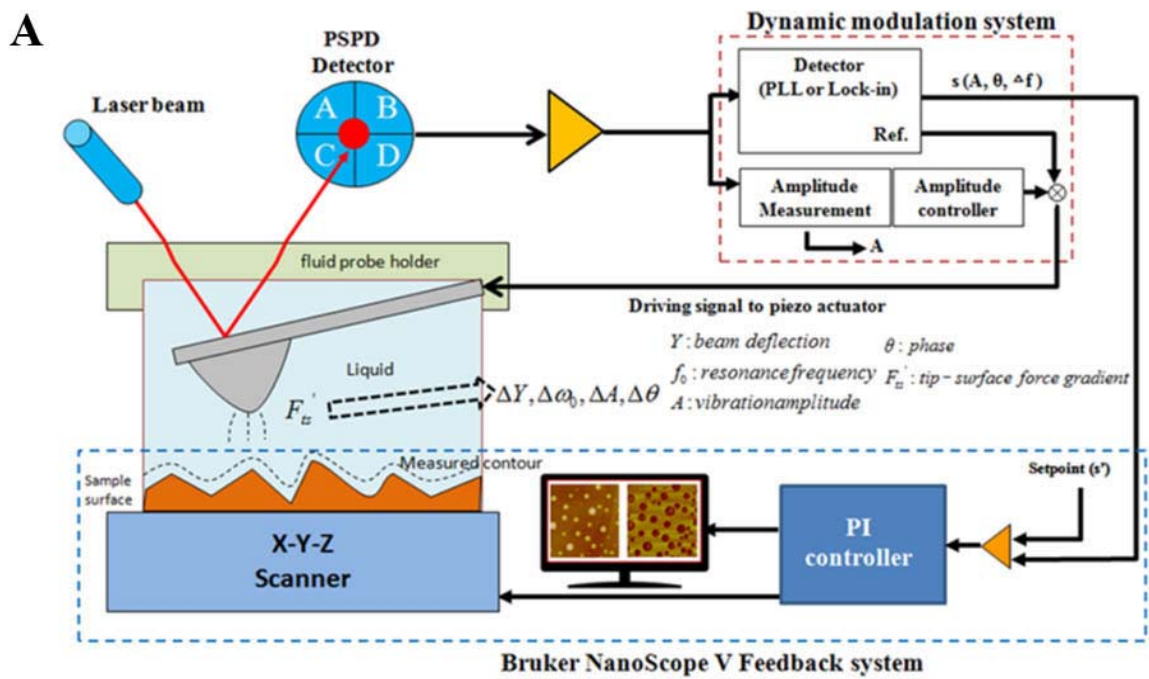

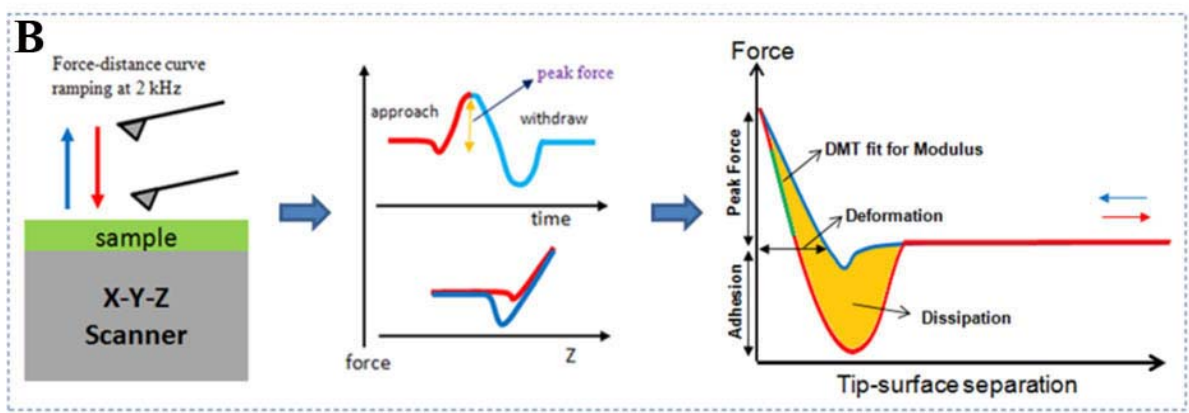

**Fig. S8.** (A) Schematic of the modified Multimode NanoScope V. (B) Illustrations of the PF mode.